%% file: Reduce_and_Boost.tex
\newif\ifuseTwoColumns

\ifuseTwoColumns
\documentclass[twocolumn]{IEEEtran}
\else
\documentclass[onecolumn,11pt,draftclsnofoot]{IEEEtran}
\usepackage{setspace}
\fi
\usepackage{amsmath}
\usepackage{amsfonts}
\usepackage{amssymb}
\usepackage{graphicx}
\usepackage{cases}
\usepackage{color}
\usepackage{subfigure}
\usepackage{multirow}
\usepackage{algorithm}
\usepackage{myalgorithmic}
\usepackage{caption}[2007/12/23]

\def\m(#1){\mathcal{#1}}
\def\b(#1){\mathbf{#1}}
\def\krank{\sigma}
\def\row(#1,#2){#1^{\textrm{row}}_#2}
\def\col(#1,#2){#1^{\textrm{col}}_#2}
\def\yl{\b(y)(\lambda)}
\def\yL{\b(y)(\Lambda)}
\def\xl{\b(x)(\lambda)}
\def\xL{\b(x)(\Lambda)}
\def\linL{\lambda\in\Lambda}

\def\bxL{\bar{\b(x)}(\Lambda)}
\def\bU{\bar{\b(U)}}

\newcommand{\rank}{\operatorname{rank} }

\newcommand{\spark}{\operatorname{Spark} }
\newcommand{\Span}{\operatorname{span} }
\def\colorname(#1){#1}
\newtheorem{theorem}{\colorname(Theorem)}
\newtheorem{lemma}{\colorname(Lemma)}

\newtheorem{proposition}{\colorname(Proposition)}

\newtheorem{definition}{\colorname(Definition)}

\def\FigImvBlock{
\begin{figure*}
\centering
\includegraphics[scale=0.7]{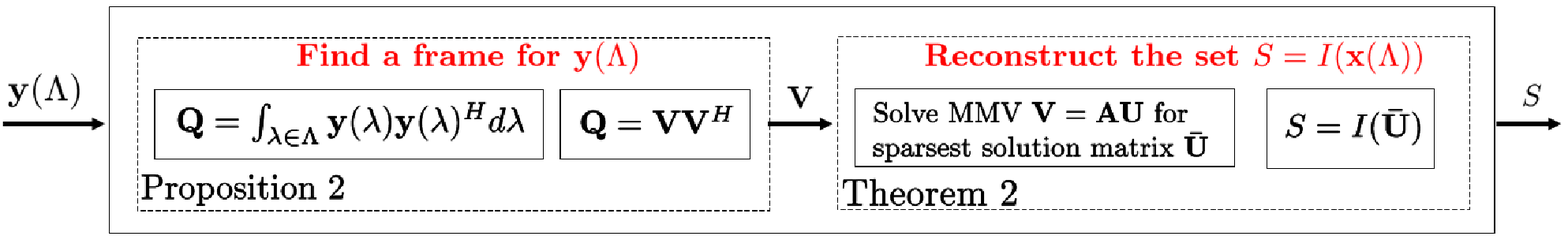}
\caption{The fundamental stages for the recovery of the non-zero
location set $S$ using only one finite-dimensional
problem.}\label{FigIMV}
\end{figure*}}

\title{
Reduce and Boost: Recovering Arbitrary Sets of Jointly Sparse
Vectors
\thanks{This work has been submitted to the IEEE for possible publication. Copyright may be transferred without notice, after which this version may no longer be accessible.}
\thanks{The authors are with the Technion---Israel Institute of
Technology, Haifa Israel. Email: moshiko@tx.technion.ac.il,
yonina@ee.technion.ac.il.}}
\author{Moshe Mishali,~\IEEEmembership{Student~Member,~IEEE}, and Yonina C.~Eldar,~\IEEEmembership{Senior~Member,~IEEE}}

\date{\today}
\usepackage[normalem]{ulem}

\begin{document}

\maketitle \IEEEpeerreviewmaketitle

\begin{abstract}
The rapid developing area of compressed sensing suggests that a
sparse vector lying in an arbitrary high dimensional space can be
accurately recovered from only a small set of non-adaptive linear
measurements. Under appropriate conditions on the measurement
matrix, the entire information about the original sparse vector is
captured in the measurements, and can be recovered using efficient
polynomial methods. The vector model has been extended both
theoretically and practically to a finite set of sparse vectors
sharing a common non-zero location set. In this paper, we treat a
broader framework in which the goal is to recover a possibly
infinite set of jointly sparse vectors. Extending existing
recovery methods to this model is difficult due to the infinite
structure of the sparse vector set. Instead, we prove that the
entire infinite set of sparse vectors can recovered by solving a
single, reduced-size finite-dimensional problem, corresponding to
recovery of a finite set of sparse vectors. We then show that the
problem can be further reduced to the basic recovery of a single
sparse vector by randomly combining the measurement vectors. Our
approach results in exact recovery of both countable and
uncountable sets as it does not rely on discretization or
heuristic techniques. To efficiently recover the single sparse
vector produced by the last reduction step, we suggest an
empirical boosting strategy that improves the recovery ability of
any given sub-optimal method for recovering a sparse vector.
Numerical experiments on random data demonstrate that when applied
to infinite sets our strategy outperforms discretization
techniques in terms of both run time and empirical recovery rate.
In the finite model, our boosting algorithm is characterized by
fast run time and superior recovery rate than known popular
methods.
\end{abstract}



\section{Introduction}

\PARstart{M}{any} signals of interest often have sparse
representations, meaning that the signal is well approximated by
only a few large coefficients in a specific basis. The traditional
strategy to capitalize on the sparsity profile is to first acquire
the signal in a high-dimensional space, and then utilize a
compression method in order to capture the dominant part of the
signal in the appropriate basis. Familiar formats like MP3 (for
audio signals) and JPEG (for images) implement this approach. The
research area of compressed sensing (CS) challenges this strategy
by suggesting that a compact representation can be acquired
directly.

The fundamentals of CS were founded in the works of Donoho
\cite{Donoho} and Cand\`{e}s \textit{et. al.} \cite{CandesRobust}.
In the basic model, referred to as a single measurement vector
(SMV), the signal is a discrete vector $\b(x)$ of high dimension.
The sensing process yields a measurement vector $\b(y)$ that is
formed by inner products with a set of sensing vectors. The key
observation is that $\b(y)$ can be relatively short and still
contain the entire information about $\b(x)$ as long as $\b(x)$ is
sparsely represented in some basis, or simply when $\b(x)$ itself
contains only a few non-zero entries. An important problem in this
context is whether the vector $\b(x)$ producing $\b(y)$ is unique
\cite{MElad}. Another well studied issue is the practical recovery
of $\b(x)$ from the compressed data $\b(y)$, which is known to be
NP-hard in general. Many sub-optimal methods have been proposed
for this problem
\cite{Donoho},\cite{CandesRobust},\cite{TroppI},\cite{TroppII},
which achieve a high recovery rate when tested on randomly
generated sparse vectors.

The SMV model has been extended to a finite set of jointly sparse
vectors having their non-zeros occurring in a common location set.
The sensing vectors are applied to each of the sparse vectors
resulting in multiple measurement vectors (MMV). This model is
well suited for problems in Magnetoencephalography, which is a
modality for imaging the brain
\cite{Gorod},\cite{MEGref1},\cite{MEGref2}. It is also found in
array processing \cite{Gorod},\cite{APref1}, nonparametric
spectrum analysis of time series \cite{SEref} and equalization of
sparse communication channels \cite{EQref1},\cite{EQref2}. The
issue of uniqueness in the MMV problem was addressed in
\cite{Cotter},\cite{Chen}, together with extensions of SMV
recovery techniques to MMV.

In this paper, we start from a broader model which consists of an
infinite set of jointly sparse vectors, termed \textit{infinite
measurement vectors} (IMV). The set may be countable or
uncountable (for example, when described over a continuous
interval). The IMV model is broader than MMV and naturally arises
in recovery problems involving analog signals, such as our earlier
work on multi-band signals \cite{MishaliSBR}. As we explain
further in the paper, the recovery of the entire infinite set of
sparse vectors in IMV models is highly complicated. A
straightforward recovery approach in this context is to consider
only a finite subset of vectors using discretization. However,
this strategy cannot guarantee perfect recovery. Instead, we
derive a reduced finite-dimensional problem from which the common
non-zero location set can be inferred exactly. This paradigm
relies on the observation that once the non-zero locations are
identified, the original recovery problem translates into a simple
linear inversion with a closed form solution.

Our first main contribution is a theoretical result showing
that for every given IMV problem there is an explicit MMV
counterpart with the same non-zero location set. This
reduction to finite dimensions is achieved without any
discretization or heuristic techniques and thus allows in
principle an exact recovery of the entire set of sparse
vectors. Other papers that treated problems involving
infinite vector sets used discretization techniques to
approximate the solution \cite{Analog2Info1} or
alternatively assumed an underlying discrete
finite-dimensional signal model \cite{Analog2Info2}. In
contrast, our approach is exact as neither the IMV model
nor the solution is discretized. Once the IMV problem is
reduced to an MMV problem, results developed in that
context can be applied.

To further improve the recovery performance both in terms of speed
and recovery rate, we develop another theoretical result allowing
to identify the non-zero locations of a given MMV model from a
sparse vector of a specific SMV system. As opposed to the IMV
reduction, our strategy here is to construct a random SMV problem
that merges the set of sparse vectors using random coefficients.
We prove that this reduction preserves the crucial information of
the non-zero location set with probability one.

Our final contribution treats the practical aspect of using a
sub-optimal technique to find the sparse vector of an SMV problem.
While examining popular SMV recovery techniques, we observed that
the recovery ability depends on the exact non-zero values and not
only on their locations. Based on this observation we argue that
it is beneficial to draw several realizations of the merged
measurement vector, by using different random combinations, until
a sparse vector is identified. These iterations are referred to as
the boost step of our method, since, empirically, each iteration
improves the overall recovery rate of the non-zero location set.
We formulate a generic algorithm, referred to as ReMBo, for the
Reduction of MMV and Boosting. The ReMBo algorithm yields
different recovery techniques for MMV based on the embedded SMV
technique.

\begin{figure}
\centering
\includegraphics[scale=0.85]{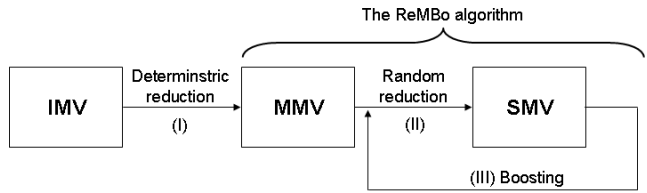}
\caption{The entire flow of the paper consists of: (I) a
deterministic reduction from IMV to MMV, (II) a random
reduction from MMV to SMV and (III) a boosting stage. The
ReMBo algorithm is a formal description of the last two
steps.}\label{FigFlow}
\end{figure}

The results presented in this paper provide a complete flow
between the recovery problem of different models.
Fig.~\ref{FigFlow} depicts the entire flow which can be initiated
from a given IMV system or an arbitrary MMV problem. Numerical
experiments demonstrate the performance advantage of methods
derived from the ReMBo algorithm over familiar MMV techniques in
terms of empirical recovery rate and run time. In addition, we
present a simulation emphasizing the advantage of the IMV
reduction over a discretization technique.

The outline of the paper is as follows. The IMV model is
introduced in Section~\ref{SecSparse}, where we also discuss
conditions for a unique solution. The deterministic reduction
method of IMV to MMV and the random reduction of MMV to SMV are
developed in Sections~\ref{SecReductionInf} and
~\ref{SecReductionMMV} respectively. The description of the ReMBo
algorithm follows in Section~\ref{SecReMBo}. Numerical experiments
are provided in Section~\ref{SecNumerical}.

\section{Infinite-Measurement-Vectors Model}\label{SecSparse}

Let $\b(A)$ be a given $m\times n$ matrix with $m<n$ and consider
the parametric linear system:
\begin{equation}\label{yAx}
\yl=\b(A)\xl,\quad\linL,
\end{equation}
where $\Lambda$ is some known set. Our goal is to recover the
unknown vector set $\xL=\{\xl\}_{\linL}$ which is referred to as
the solution set, from the measurements set $\yL=\{\yl\}_{\linL}$.
The cardinality of the parameter set $\Lambda$ is arbitrary
including both finite (single or multiple element) sets and
infinite sets (countable or uncountable). For example, $\lambda$
can be the index of a discrete set, or alternatively a variable
over a continuous interval.

Clearly, the recovery problem is not well defined unless
there is a unique solution set $\xL$ for each $\yL$.
However, the system of (\ref{yAx}) does not posses a unique
solution in general, since for every $\lambda$, (\ref{yAx})
contains less equations than unknowns. Specifically, each
$\yl$ is a vector of length $m$, while the corresponding
$\xl$ is of length $n>m$. Therefore, in order to guarantee
a unique solution an additional prior on $\xL$ must be
associated with (\ref{yAx}). Throughout this paper, we
assume the joint sparsity prior, which constrains each
$\xl$ to have only a few non-zero entries and in addition
requires that all the vectors in $\xL$ share a common
non-zero location set. The system of (\ref{yAx}) is termed
IMV when $\Lambda$ is infinite and the joint sparsity prior
is assumed. In the sequel, this prior is formally described
and is used to derive a sufficient condition for the
uniqueness of the IMV solution set.

\subsection{SMV Model}

We start by describing notation and a uniqueness result for
the special case of a single element set $\Lambda$, in
which (\ref{yAx}) is abbreviated as $\b(y)=\b(A)\b(x)$.
This corresponds to the well studied SMV model.

A vector $\b(x)$ is called $K$-sparse if it contains no more than
$K$ non-zero entries. For a given vector $\b(x)$ the support
function
\begin{equation}
I(\b(x))=\{k\,|\,\b(x)_k\neq 0\},
\end{equation}
describes the locations of the non-zero entries where $\b(x)_k$
stands for the $k$th entry of $\b(x)$. Thus, a $K$-sparse vector
$\b(x)$ conforms with a support size $|I(\b(x))|\leq K$. A
sufficient condition for the uniqueness of a $K$-sparse solution
in this setting can be stated in terms of the Kruskal-rank of a
matrix, which was first used in the seminal work of Kruskal
\cite{Kruskal}:
\begin{definition}\label{krank}
The Kruskal-rank of $\b(A)$, denoted $\sigma(\b(A))$, is the
maximal number $q$ such that every set of $q$ columns of $\b(A)$
is linearly independent.
\end{definition}
\begin{theorem}\label{SMVUniq}
If the vector $\b(\bar{x})$ is a $K$-sparse solution of
$\b(y)=\b(A)\b(x)$ and $\krank(\b(A))\geq 2K$, then $\b(\bar{x})$
is the unique $K$-sparse solution of the system.
\end{theorem}
Theorem~\ref{SMVUniq} and its proof are given in
\cite{MElad},\cite{Chen} with a slightly different notation of
$\spark(\b(A))$ instead of the Kruskal-rank.

\subsection{Uniqueness in IMV Models}

The joint sparsity prior becomes relevant when $\Lambda$
contains more than a single element. By abuse of notation,
we define the support function of a vector set as the union
over the support of each vector. Specifically,
\begin{align}
I(\xL) & =  \bigcup_{\linL} I(\xl)\\ \notag
& = \left\{1\leq k\leq n\,|\,\b(x)_k(\lambda_0)\neq 0,\textrm{ for some }\lambda_0\in\Lambda\right\}.
\end{align}
For brevity, a jointly sparse solution set $\xL$ with
$|I(\xL)|\leq K$ is also called $K$-sparse. A $K$-sparse vector
set $\xL$ implies two properties: (I) Each $\xl$ is a $K$-sparse
vector, and (II) the non-zero entries of $\xl$ are confined to a
fixed location set of size no more than $K$. The system of
(\ref{yAx}) is called MMV in the literature when the joint
sparsity prior holds over a finite set of sparse vectors
\cite{Cotter},\cite{Chen}. Similarly, we refer to the system of
(\ref{yAx}) as IMV when $\Lambda$ is an infinite set and $\bxL$ is
jointly sparse. Table~\ref{TableIMV} summarizes the models derived
from (\ref{yAx}) for different cardinalities of the set $\Lambda$.
The abbreviations used for the linear systems of MMV and IMV are
clear from the context. Evidently, the joint sparsity prior is
what distinguishes MMV and IMV models from being a set of
independent SMV systems.

\begin{table} \caption{Sparsity Models and Priors}
\begin{center}
\begin{tabular}[C]{c|c|c|l}  \hline 
Model & $\Lambda$ Cardinality & Linear System & $K$-sparsity prior \\    \hline
SMV & $1$ & $\b(y)=\b(A)\b(x)$ & $|I(\b(\bar{x}))|\leq K$\\
MMV & $d$ & $\b(Y)=\b(A)\b(X)$ & $|I(\b(\bar{X}))|\leq K$\\
IMV & Infinite & $\yL=\b(A)\xL$ & $|I(\bxL)|\leq K$\\ \hline 
\end{tabular}
\label{TableIMV}
\end{center}
\end{table}

The first property of the joint sparsity prior implies that
$\krank(\b(A))\geq 2K$ is sufficient to guarantee the uniqueness
of a $K$-sparse solution set $\bxL$, since we can consider the SMV
problem $\yl=\b(A)\xl$ for each $\lambda$ separately. Exploiting
the joint sparsity, we expect that a value of $\krank(\b(A))$ less
than $2K$ would suffice to ensure uniqueness. Extending uniqueness
results regarding MMV \cite{Cotter},\cite{Chen} leads to the
following proposition:
\begin{proposition}\label{MMVUniq}
If $\bxL$ is a $K$-sparse solution set for (\ref{yAx}), and
\begin{equation}\label{MMVUniqCond}
\krank(\b(A))\geq 2K-\left(\dim(\,\Span(\yL)\,)-1\right),
\end{equation}
then $\bxL$ is the unique $K$-sparse solution set of (\ref{yAx}).
\end{proposition}
The notation $\Span(\yL)$ is used for the subspace of minimal
dimension containing the entire vector set $\yL$. Note that
$\Span(\yL)$ is guaranteed to have finite dimension since $\yl$
has finite length. For jointly sparse solution sets,
Proposition~\ref{MMVUniq} indeed claims that the required
Kruskal-rank of $\b(A)$ can be generally lower than $2K$ of
Theorem~\ref{SMVUniq}.

\begin{proof}
The solution set $\bxL$ is $K$-sparse which implies that
$\dim(\Span(\bxL))\leq K$. It follows from the linear system of
(\ref{yAx}) that the dimension of the subspace $\Span(\yL)$ cannot
be higher than $\Span(\bxL)$, i.e. $r=\dim\{\Span(\yL)\}\leq K$.
From (\ref{MMVUniqCond}) we get that $\krank(\b(A))\geq K$.
Consequently, for each $\yl=\b(0)$ the corresponding unique
$K$-sparse vector is $\xl=\b(0)$, as the null space of $\b(A)$
cannot contain other $K$-sparse vectors. Therefore, without loss
of generality we can prove the claim for a measurement set $\yL$
with $r\geq 1$ which does not contain zero vectors.

For $r\geq 1$ there exists a finite set
$\tilde{\Lambda}=\{\lambda_i\}_{i=1}^r\subseteq\Lambda$ such that
the vector set $\b(y)(\tilde{\Lambda})$ is linearly independent.
Since $\tilde{\Lambda}$ is a finite set,
$\b(y)(\tilde{\Lambda})=\b(A)\b(x)(\tilde{\Lambda})$ is an MMV
system. According to \cite{Cotter},\cite{Chen}, the corresponding
solution set $\b(\bar{x})(\tilde{\Lambda})$ is unique under the
condition (\ref{MMVUniqCond}). Since $\yL$ does not contain zero
vectors, every vector $\yl$ belongs to some subset of $r$ linearly
independent vectors. The argument above implies the uniqueness of
the corresponding subset of $\bxL$, and consequently the
uniqueness of the entire solution set.
\end{proof}

Note that (\ref{yAx}) can be viewed as a sampling process, where
$\bxL$ is the signal, $\b(A)$ the sampling operator and $\yL$ is
the generated set of samples. In this context, the design of the
sampling operator requires to determine the number of rows in
$\b(A)$ such that the samples match a unique signal. However,
(\ref{MMVUniqCond}) cannot be used for this task since the value
of $\dim(\,\Span(\yL)\,)$ is not known a-priori. In other words,
if a matrix $\b(A)$ needs to be designed such that uniqueness is
guaranteed to every $K$-sparse solution set, including those with
$\dim\{\Span(\yL)\}=1$, then the condition (\ref{MMVUniqCond}) is
reduced to $\krank(\b(A))\leq 2K$ of Theorem~\ref{SMVUniq}.

In the remainder of the paper, we assume that a unique solution of
(\ref{yAx}) is guaranteed by either Theorem \ref{SMVUniq} or
Proposition \ref{MMVUniq}. In the next sections, we develop the
main contributions of our work which address the recovery of the
unique $K$-sparse solution set $\bxL$.

\section{Dimension Reduction for Infinite
$\Lambda$}\label{SecReductionInf}

\subsection{Optimization Viewpoint}

Before discussing the IMV model we review the optimization
viewpoint for the SMV and MMV problems.

If $\b(\bar{x})$ is the unique $K$-sparse solution of the SMV
problem $\b(y)=\b(A)\b(x)$, then it is also the unique sparsest
solution. Therefore, recovery of $\b(\bar{x})$ can be formulated
as an optimization problem \cite{Donoho}:
\begin{equation}\label{SMVP0}
\b(\bar{x})=\arg\min_\b(x) \|\b(x)\|_{\ell_0}\,\textrm{
s.t. }\,\b(y)=\b(A)\b(x),
\end{equation}
where the pseudo-norm $\ell_0$ counts the number of
non-zero entries in $\b(x)$. In our notation the objective
can be replaced by $|I(\b(x))|$. Since (\ref{SMVP0}) is
known to be NP-hard \cite{Donoho},\cite{CandesRobust},
several alternatives have been proposed in the literature.
Donoho \cite{Donoho} and Cand\`{e}s \textit{et. al.}
\cite{CandesRobust} rigorously analyze the basis pursuit
technique which uses the $\ell_1$ norm instead of the
$\ell_0$ in (\ref{SMVP0}) resulting in a tractable convex
program. Various greedy techniques to approximate the
sparsest solution have also been studied thoroughly
\cite{TroppI},\cite{TroppII},\cite{Gorod}. Empirically, all
these methods show a high recovery rate of the unique
sparsest solution when tested on random data. Analogously,
it was shown that the combinatorial problem
\begin{equation}\label{MMVP0}
\b(\bar{X})=\arg\min_\b(X) |I(\b(X))|\,\textrm{ s.t.
}\,\b(Y)=\b(A)\b(X),
\end{equation}
recovers the unique $K$-sparse solution matrix of an MMV system
\cite{Chen}. This optimization problem is also NP-hard and can be
tackled with similar efficient sub-optimal techniques
\cite{Cotter},\cite{Chen}.

Extending the optimization viewpoint to the IMV model leads to the
equivalent problem:
\begin{align}\label{IP0}
\bxL=\arg & \min_{\xL} |I(\xL)|\\ \notag & \textrm{ s.t.
}\,\yl=\b(A)\xl,\forall\linL.
\end{align}
Note that in (\ref{IP0}) there are infinitely many unknowns $\xL$,
and infinitely many equations. In contrast to the finite
formulation of both (\ref{SMVP0}) and (\ref{MMVP0}), a program of
the type (\ref{IP0}) was not analyzed in the optimization
literature. The most relevant programming structures are
semi-infinite programming \cite{SIP} and generalized semi-infinite
programming \cite{GSIP}. However, these formulations allow only
for infinite constraints while the optimization variable is
finite. This inherent intricacy of (\ref{IP0}) remains even if the
objective is relaxed by known strategies. To overcome this
difficulty, we suggest to transform (\ref{IP0}) into one of the
forms known in the literature. Specifically, we show that the
joint sparsity prior allows to convert (\ref{IP0}) into a problem
of the form (\ref{MMVP0}), in which both the variable and the
constraint set are finite.

A straightforward approach to reduce (\ref{IP0}) to a
finite-dimensional problem is to choose a finite grid
$\tilde{\Lambda}\subset\Lambda$, and then solve only for
$\b(\bar{x})(\tilde{\Lambda})$. This yields an MMV system
corresponding to the optimization problem (\ref{MMVP0}). In turn,
this program can be relaxed by any of the known CS techniques. The
final step is to approximate $\xL$ by interpolating the partial
solution set $\b(\bar{x})(\tilde{\Lambda})$. However, a
discretization approach typically results in an approximation
$\xL$ that is different from the unique solution $\bxL$. Moreover,
$\xL$ typically does not satisfy (\ref{yAx}) between the grid
points, that is for $\lambda\notin\tilde{\Lambda}$. This drawback
of discretization happens even if a brute-force method is used to
optimally find the solution set $\b(\bar{x})(\tilde{\Lambda})$ on
the grid $\tilde{\Lambda}$. Furthermore, the density of the grid
directly impacts the complexity of discretization techniques. For
these reasons, we avoid discretization and instead propose an
exact method that transforms the infinite structure of (\ref{IP0})
into a single MMV system without loosing any information. A
numerical experiment illustrating the difference between our exact
method and discretization is provided in Section~\ref{SecIMVexp}.

\subsection{Paradigm}

In order to solve (\ref{IP0}) exactly we split the problem into
two sub-problems. One is aimed at finding the support set
$S=I(\bxL)$. The other reconstructs $\bxL$ from the data $\yL$ and
the knowledge of $S$. The reason for this separation is that once
$S$ is known the linear relation of (\ref{yAx}) can be inverted
exactly. To see this, let $\b(A)_S$ denote the matrix containing
the subset of the columns of $\b(A)$ whose indices belong to $S$.
Since the solution set $\bxL$ is $K$-sparse we have that $|S|\leq
K$. In addition, from Proposition~\ref{MMVUniq},
$\krank(\b(A))\geq K$. Therefore $\b(A)_S$ consists of linearly
independent columns implying that
\begin{equation}
(\b(A)_S)^\dag\b(A)_S=\b(I),
\end{equation}
where $(\cdot)^\dag$ is the Moore-Penrose pseudo-inverse
operation. Explicitly, $(\b(A)_S)^\dag =
\left(\b(A)_S^H\b(A)_S\right)^{-1}\b(A)_S^H$ where $\b(A)_S^H$
denotes the conjugate transpose of $\b(A)_S$. Using $S$ the system
of (\ref{yAx}) can be written as
\begin{equation}\label{yAxS}
\yl = \b(A)_S \b(x)^S(\lambda),\quad\linL,
\end{equation}
where the superscript $\b(x)^S(\lambda)$ is the vector that
consists of the entries of $\xl$ in the locations $S$. Multiplying
(\ref{yAxS}) by $(\b(A)_S)^\dag$ from both sides gives
\begin{equation}\label{Reconstruct1}
\b(x)^S(\lambda) = (\b(A)_S)^\dag\b(y)(\lambda),\quad\linL.
\end{equation}
In addition, it follows from the definition of the support
set $I(\xL)$ that
\begin{equation}\label{Reconstruct2}
\b(x)_i(\lambda) = 0,\quad\forall i\notin S,\linL.
\end{equation}
Therefore (\ref{Reconstruct1})-(\ref{Reconstruct2}) allow for
exact recovery of $\bxL$ once the finite set $S$ is correctly
recovered.

\subsection{Method}

The essential part of our method is the first sub-problem that
recovers $S$ from the measurement set $\yL$. Our key observation
is that every collection of vectors spanning the subspace
$\Span(\yL)$ contains sufficient information to recover $S$
exactly, as incorporated in the following theorem:

\begin{theorem}\label{ThKey}
Suppose (\ref{yAx}) has a unique $K$-sparse solution set
$\bxL$ with $S=I(\bxL)$ and that the matrix $\b(A)_{m\times
n}$ satisfies (\ref{MMVUniqCond}). Let $\b(V)$ be a matrix
of $m$ rows such that the column span of $\b(V)$ is equal
to $\Span(\yL)$. Then, the linear system
\begin{equation}\label{VAU}
\b(V)=\b(A)\b(U)
\end{equation}
has a unique $K$-sparse solution $\bU$ and $I(\bU)=S$.
\end{theorem}

\begin{proof}
Let $r=\dim(\Span(\yL))$ and construct an $m\times r$ matrix
$\b(Y)$ by taking some set of $r$ linearly independent vectors
from $\yL$. Similarly, construct the matrix $\b(\bar{X})$ of size
$n\times r$ by taking the corresponding $r$ vectors from $\bxL$.
The proof is based on observing the linear system
\begin{equation}\label{PrThKeyYAX}
\b(Y)=\b(A)\b(X).
\end{equation}
We first prove that $\b(\bar{X})$ is the unique $K$-sparse
solution matrix of (\ref{PrThKeyYAX}) and that
$I(\b(\bar{X}))=S$. Based on this result, the matrix $\bU$
is constructed, proving the theorem.

It is easy to see that $I(\b(\bar{X}))\subseteq S$, since the
columns of $\b(\bar{X})$ are a subset of $\bxL$. This means that
$\b(\bar{X})$ is a $K$-sparse solution set of (\ref{PrThKeyYAX}).
Moreover, $\b(\bar{X})$ is also the unique $K$-sparse solution of
(\ref{PrThKeyYAX}) according to Proposition~\ref{MMVUniq}. To
conclude the claim on $\b(\bar{X})$ it remains to prove that $k\in
S$ implies $k\in I(\b(\bar{X}))$ as the opposite direction was
already proved. If $k\in S$, then for some $\lambda_0\in\Lambda$
the $k$th entry of the vector $\b(x)(\lambda_0)$ is non-zero. Now,
if $\b(x)(\lambda_0)$ is one of the columns of $\b(\bar{X})$, then
the claim follows trivially. Therefore, assume that
$\b(x)(\lambda_0)$ is not a column of $\b(\bar{X})$. We next
exploit the following lemma:
\begin{lemma}[\cite{MishaliSBR}]\label{LemIn}
For every two matrices $\b(A),\b(P)$, if $|I(\b(P))|\leq
\krank(\b(A))$ then $\rank(\b(P)) = \rank(\b(A)\b(P))$.
\end{lemma}
Clearly, Lemma~\ref{LemIn} ensures that $\rank(\b(\bar{X}))=r$. In
addition it follows from the same lemma that
$\dim\{\Span(\bxL)\}=r$. Thus, $\b(x)(\lambda_0)$ must be a
(non-trivial) linear combination of the columns of $\b(\bar{X})$.
Since the $k$th entry of $\b(x)(\lambda_0)$ is non-zero, it
implies that at least one column of $\b(\bar{X})$ has a non-zero
value in its $k$th entry, which means $k\in I(\b(\bar{X}))$.

\ifuseTwoColumns \FigImvBlock \fi

Summarizing the first step of the proof, we have that every $r$
linearly independent columns of $\yL$ form an MMV model
(\ref{PrThKeyYAX}) having a unique $K$-sparse solution matrix
$\b(\bar{X})$, such that $I(\b(\bar{X}))=S$. As the column span of
$\b(V)$ is equal to the column span of $\b(Y)$ we have that
$\rank(\b(V))=r$. Since $\b(V)$ and $\b(Y)$ have the same rank,
and $\b(Y)$ also has full column rank, we get that
$\b(V)=\b(Y)\b(R)$ for a unique matrix $\b(R)$ of $r$ linearly
independent rows. This immediately implies that
$\bU=\b(\bar{X})\b(R)$ is a solution matrix for (\ref{VAU}).
Moreover, $\bU$ is $K$-sparse, as each of its columns is a linear
combination of the columns of $\b(\bar{X})$.
Proposition~\ref{MMVUniq} implies the uniqueness of $\bU$ among
the $K$-sparse solution matrices of (\ref{VAU}).

It remains to prove that $I(\bU)=I(\b(\bar{X}))$. To simplify
notation, we write $\b(\bar{X})^i$ for the $i$th row of
$\b(\bar{X})$. Now, ${\bU}^i=\b(\bar{X})^i\b(R)$, for every $1\leq
i\leq n$. Thus, if $\b(\bar{X})^i$ is a zero row, then so is
${\bU}^i$. However, for a non-zero row $\b(\bar{X})^i$, the
corresponding row ${\bU}^i$ cannot be zero since the rows of
$\b(R)$ are linearly independent.
\end{proof}

The advantage of Theorem~\ref{ThKey} is that it allows to avoid
the infinite structure of (\ref{IP0}) and to concentrate on
finding the finite set $S$ by solving the single MMV system of
(\ref{VAU}). The additional requirement of Theorem~\ref{ThKey} is
to construct a matrix $\b(V)$ having a column span equal to
$\Span(\yL)$ (i.e. the columns of $\b(V)$ are a frame for
$\Span(\yL)$). The following proposition suggests a procedure for
creating a matrix $\b(V)$ with this property.

\begin{proposition}\label{PropFrame}
If the integral
\begin{equation}\label{MatQ}
\b(Q)
=\int_{\lambda\in\Lambda}\b(y)(\lambda)\b(y)^H(\lambda)d\lambda,
\end{equation}
exists, then every matrix $\b(V)$ satisfying
$\b(Q)=\b(V)\b(V)^H$ has a column span equal to
$\Span(\yL)$.
\end{proposition}

The existence of the integral in (\ref{MatQ}) translates into a
finite energy requirement. Specifically, for countable $\Lambda$
the integral exists if the sequence
$\{\b(y)_k(\lambda_i)\}_{i=1}^\infty$ is energy bounded in the
$\ell_2$ sense for every $1\leq k \leq m$. For uncountable
$\Lambda$, $\b(y)_k(\lambda)$ can be viewed as a function of
$\lambda$ which is required to be integrable and of bounded energy
in the $L_2$ sense for every $1\leq k \leq m$. Note that the
matrix $\b(Q)$ of (\ref{MatQ}) is positive semi-definite and thus
a decomposition $\b(Q)=\b(V)\b(V)^H$ always exists. In particular,
the columns of $\b(V)$ can be chosen as the eigenvectors of
$\b(Q)$ multiplied by the square-root of the corresponding
eigenvalues.

\begin{proof}
For finite $\Lambda$ the claim follows immediately from the fact
that every two matrices $\b(M),\b(N)$ with
$\b(M)\b(M)^H=\b(N)\b(N)^H$ have the same column space. Therefore,
it remains to extend this property to infinite $\Lambda$.

Let $r=\dim(\Span(\yL))$ and define a matrix $\b(Y)_{m\times r}$
as in the proof of Theorem~\ref{ThKey}. The columns of $\b(Y)$ are
linearly independent and thus $\b(Y)^\dag$ is well defined. Define
the vector set $\b(d)(\lambda)=\b(Y)^\dag (\yl),\linL$, where each
$\b(d)(\lambda)$ is a vector of length $r$. By construction, the
integral
\begin{equation}\label{MatE}
\int_{\lambda\in\Lambda}\b(d)(\lambda)\b(d)^H(\lambda)d\lambda=\b(Y)^\dag\b(Q)(\b(Y)^\dag)^H=\b(D)\b(D)^H
\end{equation}
exists. The last equality in (\ref{MatE}) is due to the positive
semi-definiteness of the integrand. Substituting into (\ref{MatQ})
we have that $\b(V)\b(V)^H=(\b(Y)\b(D))(\b(Y)\b(D))^H$ which
implies that the column spans of $\b(V)$ and $(\b(Y) \b(D))$ are
the same. Since the column span of $\b(Y)$ equals to $\Span(\yL)$,
$\b(d)(\Lambda)$ contains the columns of the identity matrix of
size $r\times r$, and thus $\b(D)$ is invertible. In turn, this
implies that $\Span(\b(Y))=\Span(\b(Y)\b(D))$.
\end{proof}

The computation of the matrix $\b(Q)$ depends on the underlying
application. In \cite{MishaliSBR} we considered this approach for
the reconstruction of an analog multi-band signal from point-wise
samples. This class of signals are sparsely represented in the
frequency domain as their Fourier transform is restricted to
several disjoint intervals. Imposing a blind constraint, namely
that both sampling and reconstruction are carried out without
knowledge of the band locations, yields an IMV system that depends
on a continuous frequency parameter. As described in
\cite{MishaliSBR}, in this application $\b(Q)$ can be computed by
evaluating correlations between the sampling sequences in the time
domain. The existence of the integral in (\ref{MatQ}) corresponds
to the basic requirement that the point-wise sampling process
produces bounded energy sampling sequences.

\ifuseTwoColumns \else \FigImvBlock \fi

Fig.~\ref{FigIMV} summarizes the reduction steps that follow from
Theorem~\ref{ThKey} and Proposition~\ref{PropFrame}. The flow of
Fig.~\ref{FigIMV} was first presented and proved in our earlier
work \cite{MishaliSBR}. The version we provide here has several
improvements over the preliminary one of \cite{MishaliSBR}. First,
the flow is now divided into two independent logical stages and
the purpose of each step is highlighted. Second, each stage has a
stand-alone proof as opposed to the technique used in
\cite{MishaliSBR} to prove the entire scheme at once.
Mathematically, this separation allows us to remove the
restriction imposed in \cite{MishaliSBR} on $\b(V)$ to have only
orthogonal columns. Moreover, each block can be replaced by
another set of operations having an equivalent functionality. In
particular, the computation of the matrix $\b(Q)$ of
Proposition~\ref{PropFrame} can be avoided if other methods are
employed for the construction of a frame $\b(V)$ for $\Span(\yL)$.

\section{Dimension Reduction for Finite
$\Lambda$}\label{SecReductionMMV}

\subsection{Objective}

We now address the finite case of an MMV system
\begin{equation}\label{MMVYAX}
\b(Y)=\b(A)\b(X),
\end{equation}
with $\b(A)$ an $m\times n$ rectangular matrix as before.
Following the convention of Table~\ref{TableIMV}, $\b(Y)$ is an
$m\times d$ matrix, and the dimensions of $\b(X)$ are $n\times d$.
We assume that a unique $K$-sparse solution matrix $\b(\bar{X})$
with no more than $K$ non-identical zero rows exists. The unique
solution $\b(\bar{X})$ can be found by the optimization problem
(\ref{MMVP0}), which has known relaxations to tractable
techniques. Our goal in this section is to rely on ideas developed
in the context of the IMV model in order to reduce the dimension
of the optimization variable of (\ref{MMVP0}) before performing
any relaxation. Note that the MMV system (\ref{MMVYAX}) is
arbitrary and the results developed in the sequel do not assume a
preceding stage of reduction from IMV.

Applying the same paradigm of the infinite scenario, we aim
to recover the support set $S=I(\b(\bar{X}))$. This set
contains the crucial information in the sense that once $S$
is recovered the solution is obtained by
(\ref{Reconstruct1})-(\ref{Reconstruct2}), namely by
inverting the relevant columns of $\b(A)$. An immediate
corollary of Theorem~\ref{ThKey} is that if $\b(Y)$ does
not have full column rank, then (\ref{MMVYAX}) can be
reduced by taking an appropriate column subset of $\b(Y)$.
However, we wish to improve on this trivial result.
Specifically, we intend to find the support set $S$ from a
single SMV optimization of the form (\ref{SMVP0}). Such a
reduction method is beneficial as the dimensions of the
unknown variable in (\ref{SMVP0}) is $n$ while in
(\ref{MMVP0}) it is $nd$.

\subsection{Method}

Our approach is to randomly merge the columns of $\b(Y)$ into a
single vector $\b(y)$. We then show that the set $S$ can be
extracted from the random SMV $\b(y)=\b(A)\b(x)$. In order to
derive this result rigorously we rely on the following definition
from probability and measure theory
\cite{ProbBook},\cite{MeasBook}:

\begin{definition}\label{DefAbsCont} 
A probability distribution $\m(P)$ is called \textit{absolutely
continuous} if every event of measure zero occurs with probability
zero.
\end{definition}

A distribution is absolutely continuous if and only if it can be
represented as an integral over an
integrable density function \cite{ProbBook},\cite{MeasBook}. 
For example, Gaussian and uniform distributions have an explicit
density function that is integrable and thus both are absolutely
continuous. Conversely, discrete and other singular distributions
are not absolutely continuous. The following theorem exploits this
property to reduce (\ref{MMVP0}) into (\ref{SMVP0}):

\begin{theorem}\label{ThMerge}
Let $\b(\bar{X})$ be the unique $K$-sparse solution matrix of
(\ref{MMVYAX}) with $\krank(\b(A))\geq 2K$. In addition, let
$\b(a)$ be a random vector of length $d$ with an absolutely
continuous distribution and define the random vectors
$\b(y)=\b(Y)\b(a)$ and $\b(\bar{x})=\b(\bar{X})\b(a)$. Then, for
the random SMV system $\b(y)=\b(A)\b(x)$ we have:
\begin{enumerate}
\item For every realization of $\b(a)$, the vector
    $\b(\bar{x})$ is the unique $K$-sparse solution of the
    SMV. \item $I(\b(\bar{x}))=I(\b(\bar{X}))$ with
    probability one.
\end{enumerate}
\end{theorem}

\begin{proof}
For every realization of $\b(a)$, the vector $\b(\bar{x})$ is a
linear combination of the jointly $K$-sparse columns of
$\b(\bar{X})$, and thus $\b(\bar{x})$ is  $K$-sparse. It is easy
to see that $\b(\bar{x})$ satisfies the SMV system and that
Theorem~\ref{SMVUniq} implies its uniqueness among the $K$-sparse
vectors.

Denote $S=I(\b(\bar{X}))$ and observe that the previous argument
implies that $I(\b(\bar{x}))\subseteq S$ for every realization of
$\b(a)$. Therefore, it remains to prove that the event
$I(\b(\bar{x}))=S$ occurs with probability one. Expressing this
event in terms of the rows of $\b(\bar{X})$ gives
\begin{align}\label{ProofThMerge1}
\textrm{Prob}\left\{I(\b(\bar{x})) = S\right\}
 & =\textrm{Prob}\left\{\b(a)\notin\mathcal{N}\left(\b(\bar{X})^i\right),\quad\forall
i\in S\right\}\\
\notag & = 1 - \textrm{Prob}\left\{\b(a)\in\bigcup_{i\in S}
\mathcal{N}\left(\b(\bar{X})^i\right)\right\},
\end{align}
where $\b(\bar{X})^i$ denotes the $i$th row of $\b(\bar{X})$, and
$\mathcal{N}\left(\b(\bar{X})^i\right)=\{\b(v)\,|\,\b(\bar{X})^i
\b(v)=\b(0)\}$ is the null space of that row. Now, for every $i\in
S$, the row $\b(\bar{X})^i$ is not identically zero, and thus the
dimension of $\mathcal{N}\left(\b(\bar{X})^i\right)$ is $d-1$. In
other words, for every $i\in S$, we have that
$\mathcal{N}\left(\b(\bar{X})^i\right)$ has a zero measure in the
underlying sample space of $\b(a)$, which can be either
$\mathbb{R}^d$ or $\mathbb{C}^d$. The union in
(\ref{ProofThMerge1}) is over the finite set $S$ and thus has also
a measure zero. The absolutely continuity of the distribution of
$\b(a)$ concludes the proof.
\end{proof}

The randomness of $\b(a)$ plays a main role in the
reduction method suggested by Theorem~\ref{ThMerge}. In
fact, random merging is a best choice in the sense that for
every deterministic linear merging there are infinite
counterexamples in which the merging process would fail to
preserve the support set $S$. For example, a simple
summation over the columns of $\b(Y)$ may fail if the
non-zero values in a single row of $\b(\bar{X})$ sum to
zero. In contrast, Theorem~\ref{ThMerge} ensures that for
every given MMV system and with probability one, the random
reduction yields an SMV with the same non-zero location
set.

The result of Theorem~\ref{ThMerge} resembles a result of
\cite{DCS}, in which the authors suggested merging the columns of
$\b(Y)$ using an ordinary summation. The non-zero locations were
then estimated using a one step greedy algorithm (OSGA). It was
shown that if the entries of $\b(\bar{X})$ are random, drawn
independently from a Gaussian distribution, then the set $S$ can
be recovered by OSGA with probability approaching one as long as
$d\rightarrow \infty$, that is when the number of columns in each
of the matrices $\b(Y),\b(\bar{X})$ is taken to infinity. In
contrast, our method does not assume a stochastic prior on the
solution set $\b(\bar{X})$. Moreover, Theorem~\ref{ThMerge} holds
with probability one for arbitrary finite and fixed values of $d$.

\section{The ReMBo Algorithm}\label{SecReMBo}

Theorem~\ref{ThMerge} paves the way to a new class of MMV
techniques based on reduction to an SMV. In this approach,
the measurement matrix $\b(Y)$ is first transformed into a
single vector $\b(y)$ by drawing a realization of $\b(a)$
from some absolutely continuous distribution. Then, an SMV
problem of the type (\ref{SMVP0}) is solved in order to
find the support set $S$. Finally, the recovery of
$\b(\bar{X})$ is carried out by inverting the matrix
$\b(A)_S$ as in (\ref{Reconstruct1})-(\ref{Reconstruct2}).

Since (\ref{SMVP0}) is NP-hard it is not solved explicitly in
practice. Instead, many efficient sub-optimal techniques have been
proposed in the literature that are designed to be tractable but
no longer guarantee a recovery of the unique sparsest solution.
Interestingly, we have discovered that repeating the reduction
process of the previous section with different realizations of
$\b(a)$ is advantageous due to the following empirical behavior of
these sub-optimal techniques. Consider two $K$-sparse vectors
$\b(\bar{x}),\b(\tilde{x})$ having the same non-zero locations but
with different values. Denote by $\mathcal{S}$ an SMV technique
which is used to recover $\b(\bar{x}),\b(\tilde{x})$ from the
measurement vectors $\b(A)\b(\bar{x}),\b(A)\b(\tilde{x})$
respectively. Empirically, we observed that $\mathcal{S}$ may
recover one of the vectors $\b(\bar{x}),\b(\tilde{x})$ while
failing to recover the other, even though their non-zero locations
are the same. As far as we are aware, this behavior was not
studied thoroughly yet in the literature. In fact, Monte-Carlo
simulations that are typically conducted in the evaluation of CS
techniques may imply a converse conclusion. For example,
Cand\`{e}s \textit{et. al.} \cite{CandesRobust} analyzed the basis
pursuit method for SMV when $\b(A)$ is a row subset of the
discrete time Fourier matrix. A footnote in the simulation section
points out that the observed behavior seems to be independent of
the exact distribution of which the non-zero entries are drawn
from. This remark was also validated by other papers that
conducted similar experiments. The conjecture that Monte-Carlo
simulations are insensitive to distribution of the non-zero values
appears to be true. Nevertheless, it is beneficial for a given SMV
system to apply $\mathcal{S}$ on both measurement vectors
$\b(A)\b(\bar{x}),\b(A)\b(\tilde{x})$. Once the crucial
information of the non-zero locations is recovered, the final step
of inverting $\b(A)_S$ leads to the correct solution of both
$\b(\bar{x}),\b(\tilde{x})$.

The ReMBo algorithm, outlined in Algorithm~\ref{Alg1}, makes use
of the reduction method and also capitalizes on the empirical
behavior discussed above. In steps \ref{StepRed1}-\ref{StepRed2},
the MMV system is reduced into an SMV and solved using a given SMV
technique $\mathcal{S}$. These steps produce a sub-optimal
solution $\b(\hat{x})$, which is examined in step
\ref{StepExamine}. If $\b(\hat{x})$ is not sparse enough or is not
well aligned with the measurements, then the reduction steps are
repeated with another draw of the random vector $\b(a)$. We term
these additional iterations the boosting step of the algorithm.
Theorem~\ref{ThMerge} ensures that each of the different SMV
systems of step \ref{StepSMV} has a sparse solution that preserves
the required support set $S$ with probability one. The iterations
improve the chances to recover $S$ by changing the non-zero values
of the sparse solutions. Note that if the number of iterations
exceed the pre-determined parameter $\textsf{MaxIters}$, then the
algorithm is terminated. The content of the \textsf{flag} variable
indicates whether $\b(\hat{X})$ represents a valid solution. If
\textsf{flag=false}, then we may solve the MMV system by any other
method.

\input{ReMBo_alg.tex}

In general, CS techniques can be clustered into two groups. Those
of the first group search for the sparsest feasible solution. The
other group contains approximation methods that fix the sparsity
to a user-defined value and determine a solution in this set that
is best aligned with the data. For example, basis pursuit
\cite{BPref2} belongs to the first group, while matching pursuit
\cite{MPref} with a fixed number of iterations belongs to the
second group. The ReMBo algorithm can be tuned to prefer either
feasibility or sparsity according to user preference by selecting
appropriate values for the parameters $K,\epsilon$. However, it is
recommended to avoid an approximation technique of the second
group when constraining only $K$ to a desired value. The reason is
that such a method makes the condition of step \ref{StepExamine}
always true, and thus no boosting will occur.

We now compare the behavior of ReMBo with standard MMV
techniques in terms of computational complexity and
recovery rate. Clearly, the complexity of SMV is lower due
to the reduced number of unknowns. The reduction method
itself is no more than one matrix multiplication which in
practice is a negligible portion of the overall run time in
typical CS techniques. Performance of different algorithms
can also be evaluated by measuring the empirical recovery
rate in a set of random tests
\cite{Donoho},\cite{CandesRobust},\cite{Cotter},\cite{Chen}.
As we detail in the following section, for some parameters
choices a single reduction iteration achieves an overall
recovery rate that is higher than applying a direct MMV
technique. For other parameter selections, a single
iteration is not sufficient and boosting is required to
increase the recovery rate of a ReMBo technique beyond that
of a standard MMV. The results indicate that ReMBo based
techniques are comparably fast even when boosting is
employed.

\section{Numerical experiments}\label{SecNumerical}

In this section we begin by evaluating the reduction and
boosting approach for MMV systems. The behavior of the
ReMBo algorithm is demonstrated when the produced SMV is
solved using a sub-optimal method. Two representative MMV
techniques are derived from Algorithm~\ref{Alg1} and
compared with other popular MMV techniques. We then present
an experiment that demonstrates the benefits of the IMV
reduction flow over a discretization technique.

\ifuseTwoColumns \input{list_techniques.tex} \fi

\subsection{Evaluating ReMBo}

We choose $m=20,n=30,d=5$ for the dimensions of (\ref{MMVYAX}).
The following steps are repeated 500 times for each MMV technique:
\begin{enumerate}
    \item A real-valued matrix $\b(A)$ of size $20 \times 30$
        is constructed by drawing each of its entries
        independently from a Gaussian distribution with zero
        mean and variance one.

    \item For each value of $1\leq K\leq 20$ we construct a
        $K$-sparse real-valued solution matrix $\b(\bar{X})_K$
        of size $30\times 5$. The non-zero values of
        $\b(\bar{X})_K$ are also drawn from a Gaussian
        distribution in the same way described before.

    \item The MMV technique that is being tested is executed
        in order to recover each $\b(\bar{X})_K$ from the
        measurement data $\b(A)\b(\bar{X})_K$. For ReMBo
        techniques, $\m(P)$ is an i.i.d. uniform
        distribution in
        $[-1,1]^d$.

    \item A correct solution is announced
    if $\b(\bar{X})_K$ is recovered exactly up to machine
    precision.
\end{enumerate}
The empirical recovery rate for each value of $K$ is calculated as
the percentage of correct solutions. We also collected run time
data in order to qualitatively compare between the time complexity
of the tested techniques. Rigorous complexity analysis and
comparison are beyond the scope of this paper. Note that the
selection of real-valued matrices is not mandatory and the results
are also valid for complex values. However, we stick to the
real-valued setting as it reproduces the setup of
\cite{Cotter},\cite{Chen}. In addition, the same empirical
recovery rate is noticed when the non-zero entries of
$\b(\bar{X})_K$ are drawn from a non-Gaussian distribution (e.g.
uniform distribution). This behavior strengthens the conjecture
that Monte-Carlo analysis is insensitive to the specific
distribution of the non-zero values.

\ifuseTwoColumns \else \input{list_techniques.tex} \fi

To simplify the presentation of the results,
Table~\ref{TableTechniques} lists the techniques that are
used throughout the experiments. Short labels are used to
denote each of the techniques. The notation
$\m(R)_{\ell_p}(\b(X))$ stands for a vector of length $n$
such that its $i$th entry is equal to the $\ell_p$ norm of
the $i$th row of $\b(X)$. In the sequel we denote the
$\textsf{MaxIters}$ parameter of ReMBo based techniques in
brackets, for example ReMBo-BP[1]. A default value of
$\textsf{MaxIters}=\rank(\b(Y))$ is used if the brackets
are omitted. This selection represents an intuitive choice,
since after $\rank(\b(Y))$ iterations, step \ref{Step_y} of
Algorithm~\ref{Alg1} produces a vector $\b(y)$ that is
linearly dependent in the realizations of the previous
iterations. This intuition is discussed later in the
results.

Note that there is a difference in deciding on a correct solution
for SMV and MMV. In the latter, a solution is considered correct
only when all the vectors in the matrix are recovered
successfully, while in SMV a recovery of a single vector is
required. Nevertheless, as both problems amount to recovering the
finite support set, we plot the recovery rate curves of SMV and
MMV techniques on the same scale. An alternative approach would be
to adjust the SMV recovery curve so that it represents the overall
success rate when the SMV technique is applied to each of the
columns separately. Adjusting the results according to this
approach will only intensify the improved recovery rate of ReMBo
based techniques.

\subsection{Results}

\begin{figure}
\centering
\includegraphics[scale=0.7]{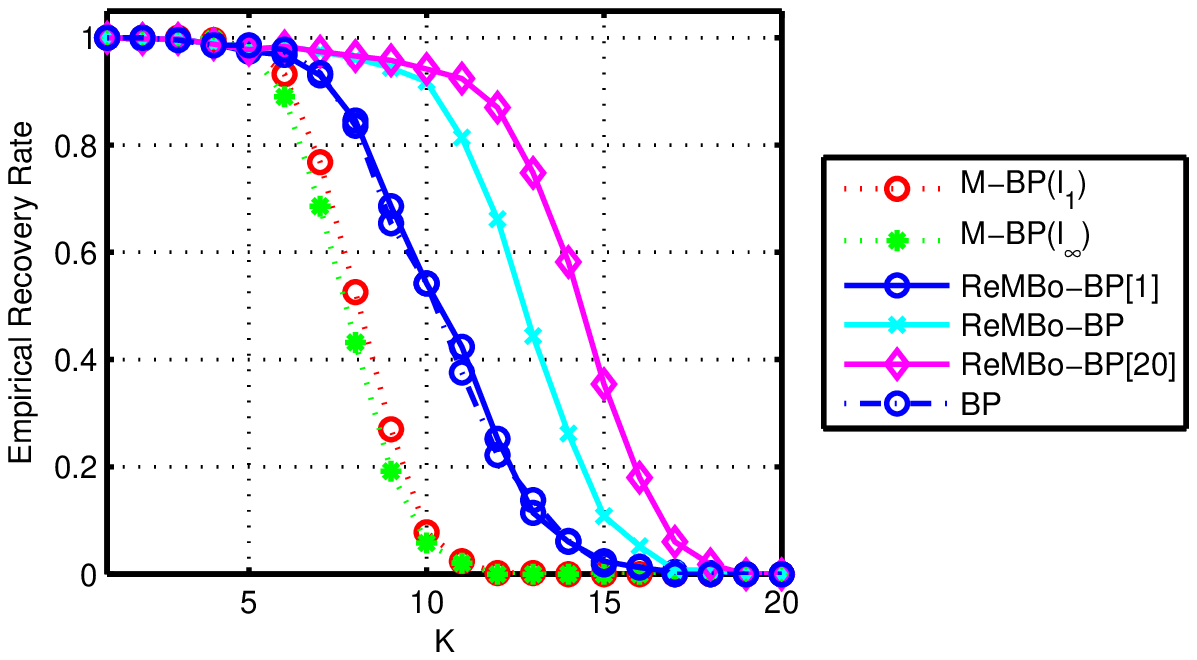}
\caption{Comparison of MMV techniques based on convex relaxations. The ReMBo techniques are in solid lines. As expected, the recovery curves of
ReMBo-BP[1] and BP coincide. }\label{Fig1}
\end{figure}

In Fig.~\ref{Fig1} we compare between MMV techniques based on
convex relaxation of (\ref{MMVP0}). For reference we also draw the
recovery rate of BP on a single measurement column. It is seen
that both M-BP($\ell_1$) and M-BP($\ell_\infty)$ suffer from a
decreased recovery rate with respect to BP. In contrast, the
recovery rate of ReMBo-BP improves on BP due to the boosting
effect. In addition, as revealed from Fig.~\ref{FigRuntimes} the
average run time of ReMBo-BP is also lower than the run time of
either M-BP($\ell_1$) and M-BP($\ell_\infty)$. Clearly, this
simulation shows that besides the theoretical interest in the
special convex relaxation of M-BP-$\ell_1$ and M-BP-$\ell_\infty$,
in this example these method do not offer a practical benefit.
Furthermore, the M-BP techniques require the selection of a row
norm besides the standard selection of $\ell_1$ norm for the final
column vector. The reduction method allows to avoid this ambiguous
selection by first transforming to an SMV problem.

Matching pursuit (including its variations) and FOCUSS are both
greedy methods that construct the set $S$ iteratively. These
techniques are typically faster than basis pursuit based methods
as seen in Fig.~\ref{FigRuntimes}. In addition, extending the SMV
version of these techniques into MMV is immediate. As opposed to
convex relaxation methods, these approaches demonstrate an
improved recovery rate when a joint sparsity prior is introduced.
This behavior is depicted in Fig.~\ref{Fig2}. A comparison of
these methods with ReMBo techniques is shown in Fig.~\ref{Fig3}.
It is seen that ReMBo-OMP outperforms M-OMP and M-FOCUSS over the
range $1\leq K \leq 13$. Specifically, in the intermediate range
$10\leq K \leq 13$ it reaches a recovery rate that is
approximately 10\% higher than the maximal recovery rate of the
non-ReMBo techniques. In addition, the run time of the ReMBo-OMP
is not far from the direct greedy approaches as seen from
Fig.~\ref{FigRuntimes}.

\ifuseTwoColumns
\def\Fig2Scale{0.5}
\else
\def\Fig2Scale{0.7}
\fi
\begin{figure}
\centering \mbox {
\subfigure[OMP]{\includegraphics[scale=\Fig2Scale]{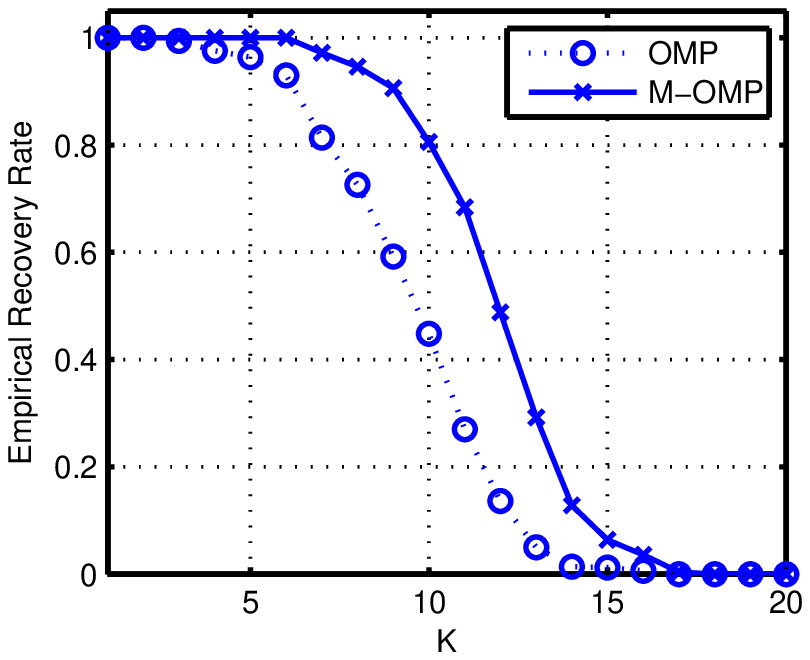}}
\subfigure[FOCUSS]{\includegraphics[scale=\Fig2Scale]{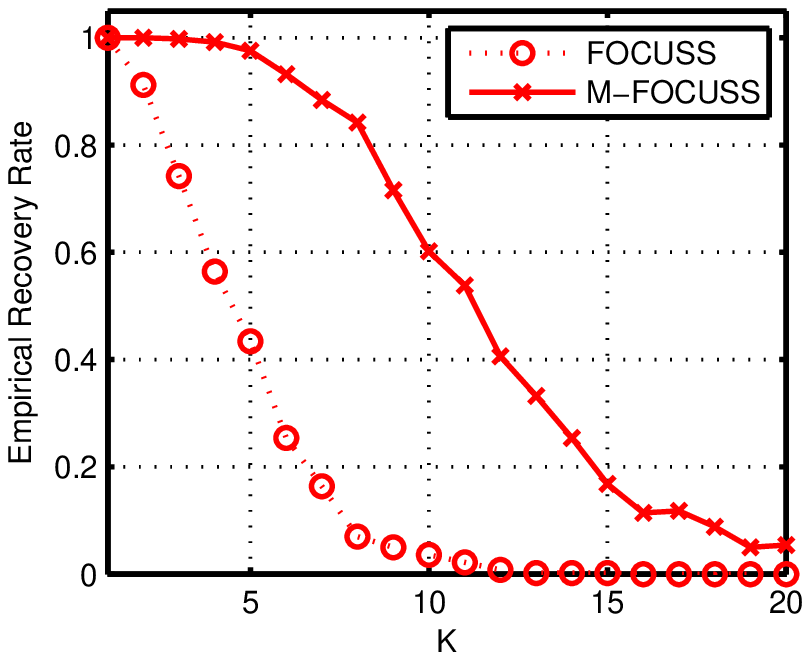}}}
\caption{The recovery rate of sequential selection
techniques is demonstrated for MMV and SMV with the same
number of non-zero entries per solution vector. The stopping criteria for OMP is based on the residual. The FOCUSS algorithm is designed to produce a $K$-sparse approximation of the solution (for this reason a ReMBo-FOCUSS method is not tested as it cannot exploit the boosting strategy).} \label{Fig2}
\end{figure}

\begin{figure} \centering
\includegraphics[scale=0.7]{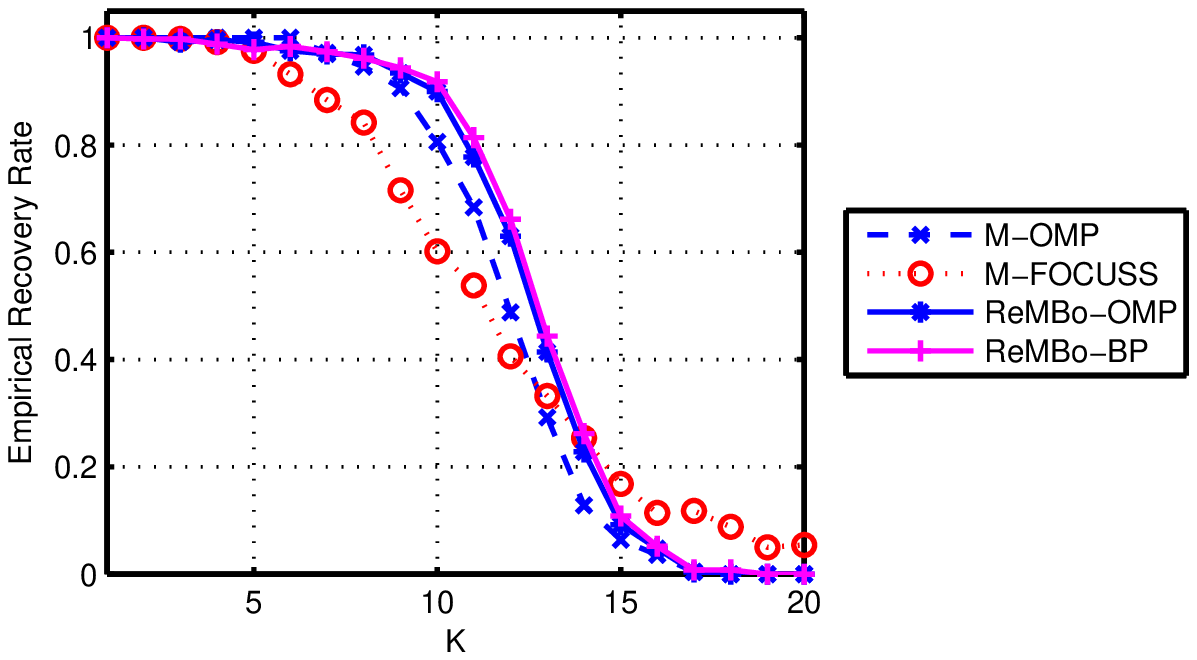}
\caption{A comparison between popular MMV techniques and
ReMBo derived methods.}\label{Fig3}
\end{figure}

In order to emphasize the impact of iterations, Fig.~\ref{Fig4}
depicts the recovery rate of ReMBo-BP and ReMBo-OMP for different
values of \textsf{MaxIters}. The recovery rate at $K=10$ is of
special interest as according to Theorem~\ref{ThMerge}
$\krank(\b(A))\geq 2K$ is required\footnote{According to
\cite{Donoho},\cite{CandesRobust}, a matrix with random entries
has a full column rank and a full Kruskal rank with an
overwhelming probability. In our setup the maximal value of
$\krank(\b(A))$ is $m=20$. Empirically, it was also noticed that
$\rank(\b(Y))=5$ in all generated measurements.} to ensure that
the random instances of SMV preserve the set $S$. For example, a
single iteration of ReMBo-BP achieves a recovery rate of 54\%,
while two and five iterations improve the recovery rate to 74\%
and 91\% respectively. A higher number of iterations results in a
minor improvement conforming with our intuitive default selection
of $\textsf{MaxIters}=\rank(\b(Y))$. However, the condition of
$K\leq 10$ is only sufficient and empirical recovery is allowed to
some extent even for $K>10$. This behavior is common to all the
techniques tested here as shown in Figs.~\ref{Fig1}-\ref{Fig4}. In
this range of $K>10$, repeating the reduction process for more
than $\rank(\b(Y))$ can be beneficial. For example, ReMBo-BP[20]
yields a recovery rate of 56\% for $K=14$ instead of 25\% when
allowing only \textsf{MaxIters}=5.

\ifuseTwoColumns
\def\Fig4Scale{0.5}
\else
\def\Fig4Scale{0.7}
\fi
\begin{figure}
\centering \mbox {
\subfigure[ReMBo-BP]{\includegraphics[scale=\Fig4Scale]{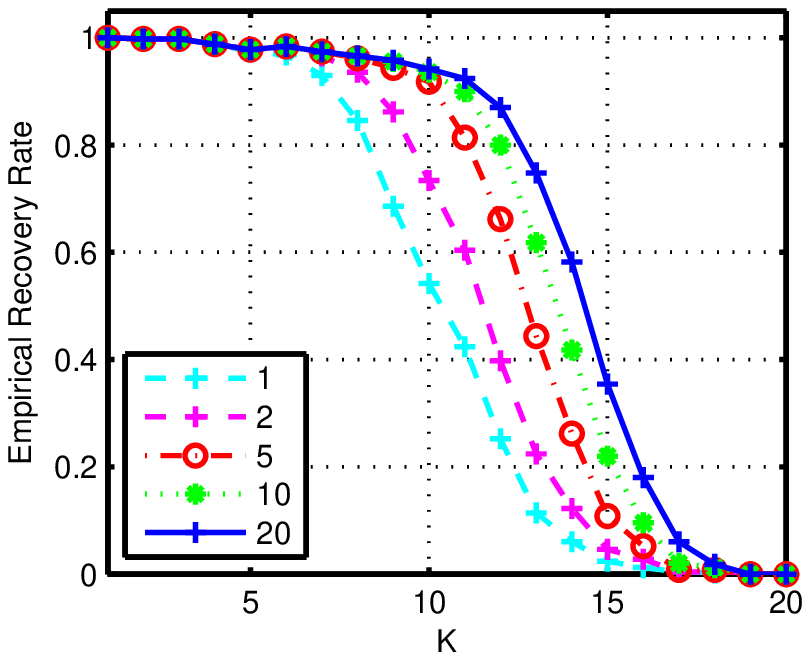}}
\subfigure[ReMBo-OMP]{\includegraphics[scale=\Fig4Scale]{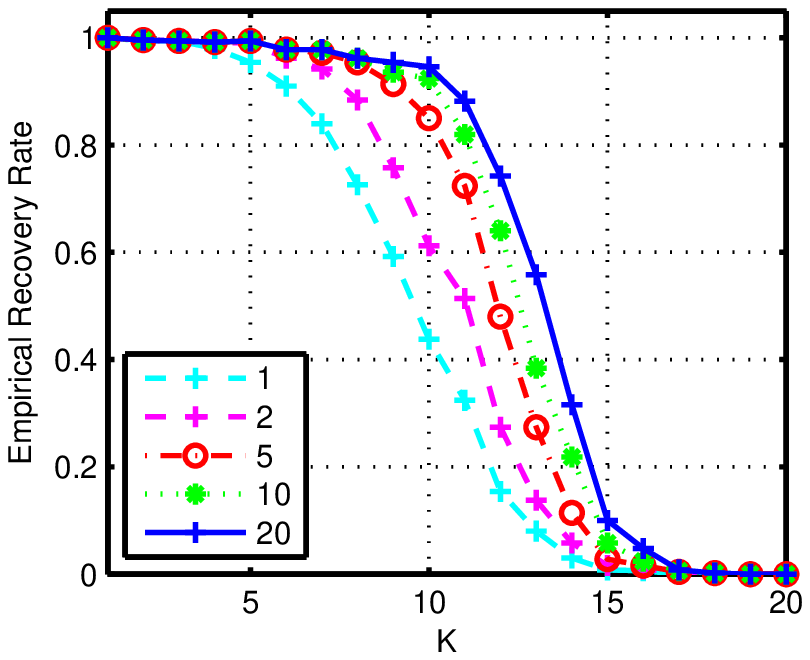}}}
\caption{The impact of boosting iterations for various selections of $\textsf{MaxIters}$.} \label{Fig4}
\end{figure}

\begin{figure}
\centering
\includegraphics[scale=0.7]{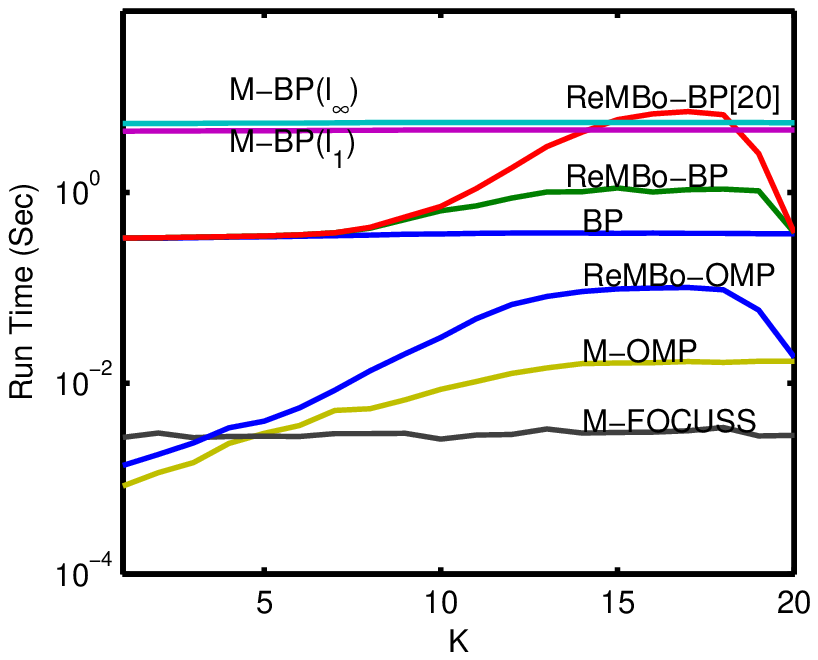}
\caption{Average run time of various MMV
techniques.}\label{FigRuntimes}
\end{figure}

\subsection{IMV Reduction vs. Discretization}\label{SecIMVexp}

We now extend the previous setup in order to simulate an IMV model
by letting $d=10000$. To discretize the IMV system, $g$ evenly
spaced columns of $\b(Y)$ are chosen resulting in an MMV system
whose sparsest solution is searched, where $1\leq g \leq 200$.
Since the non-zero values are drawn randomly, interpolation of the
missing columns is no useful in this setting. Instead, we consider
an approximation of the non-zero location set $S$ by taking the
support of the solution matrix on the chosen grid. Finally, the
entire solution set is recovered by
(\ref{Reconstruct1})-(\ref{Reconstruct2}). In order to capitalize
on the difference between the IMV reduction flow of
Fig.~\ref{FigFlow} and this discretization technique, we consider
$K$-sparse solution matrices $\b(\bar{X})_K$ such that each
non-zero row of $\b(\bar{X})_K$ has only a few non-zero entries
(e.g. up to 150 non-zero values). For a fair comparison, the M-OMP
technique is used for the recovery of $\b(\bar{X})_K$ in both
methods.

The empirical recovery rate for several values of $g$ is shown in
Fig.~\ref{FigIMVresults}. It is evident that a discretization
technique of this type requires a grid of $g=200$ to approach a
reasonable recovery rate, which is still below the recovery rate
of the IMV flow. In order to explain the superior performance of
the IMV flow we plot a typical structure of a solution set in
Fig.~\ref{FigDisc}. It is clear that discretization may fail as it
does not capture the entire information of the solution set. In
contrast, our approach preserves the necessary information
required for perfect reconstruction of the solution set, namely
the non-zero location set. Furthermore, comparing the average run
time of both approaches reveals that IMV is even faster than
discretization having a similar recovery rate. Note that the
density of the grid influences the run time of discretization
methods. In the example above of $g=200$, discretization yields an
MMV system with 200 columns. The IMV flow does not have this
drawback, as it follows from Lemma~\ref{LemIn} that the matrix
$\b(V)$ can be chosen such that it consists of no more than $K\leq
20$ columns.

\ifuseTwoColumns
\def\FigIMVScale{0.5}
\else
\def\FigIMVScale{0.7}
\fi
\begin{figure}
\centering \mbox {
\subfigure[]{\includegraphics[scale=\FigIMVScale]{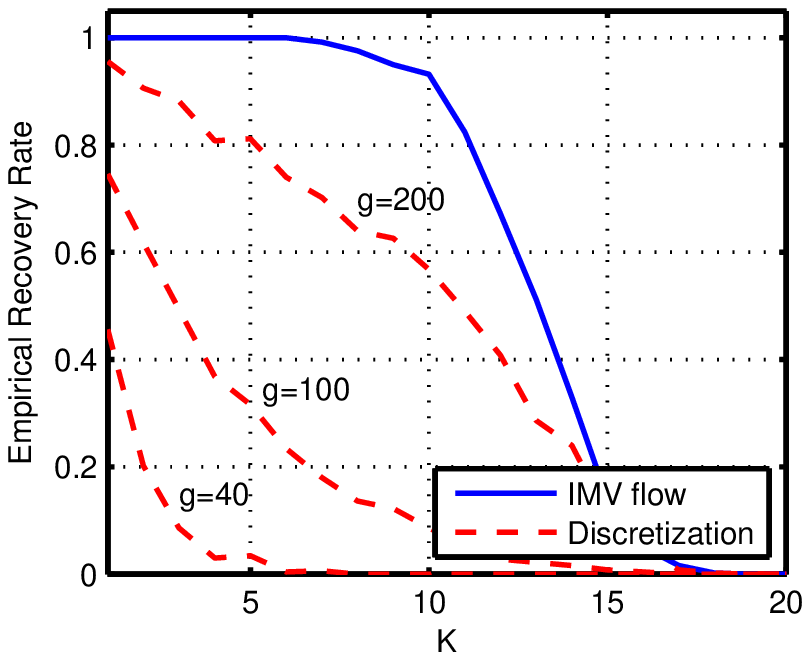}}
\subfigure[]{\includegraphics[scale=\FigIMVScale]{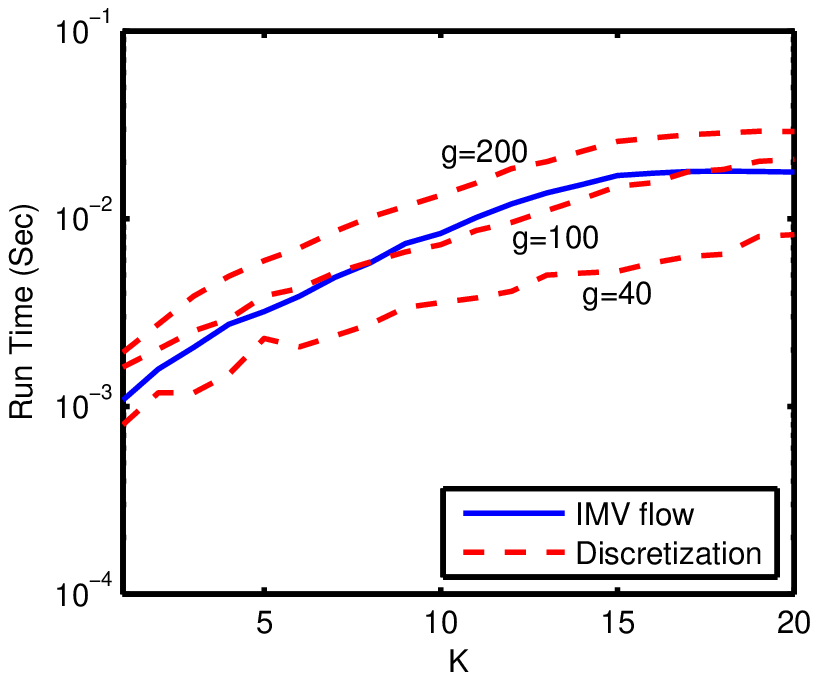}}}
\caption{A comparison of (a) the recovery rate and (b) the average run time between the IMV flow and discretization.} \label{FigIMVresults}
\end{figure}

\begin{figure}
\centering
\includegraphics[scale=1]{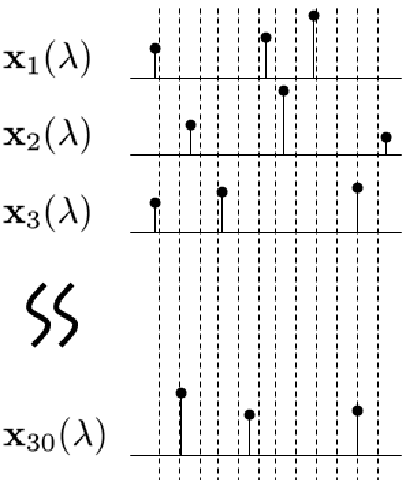}
\caption{A typical structure of the solution set and a grid
selection. The grid cannot be synchronized with the
non-zero locations. In this example, discretization
technique would fail to reconstruct $\b(x)_2(\lambda)$
whereas the IMV flow guarantees an exact recovery of the
support set $S$.}\label{FigDisc}
\end{figure}

\section{Conclusions}

The essence of the reduction theorems developed in this paper is
that the recovery of an arbitrary number of jointly sparse vectors
amounts to solving a single sparse vector of an SMV. This result
applies to the finite case of MMV and to the broader model of IMV
which we introduced here. The key observation used in our
developments is that the non-zero location set is the crucial
information for the exact recovery of the entire solution set. We
prove that this set can be recovered from a low dimensional
problem rather than directly from the given high dimensional
system.

The explicit recovery problem of sparse vectors is a difficult
combinatorial optimization program. Various methods to approximate
the sparse solution of a given program have been previously
proposed. However, to the best of our knowledge, a direct
simplification of the explicit combinatorial formulation, in the
way described here, was not studied so far. Furthermore, in a
typical CS setting the sensing process involves randomness while
the reconstruction is deterministic. The reduction method for MMV
shows that randomness can also be beneficial in the reconstruction
stage. In addition, popular recovery techniques have a fixed
performance in terms of run time and recovery rate. In contrast,
the ReMBo algorithm is tunable as it allows to trade the run time
by the overall recovery rate. The simulations conducted on several
ReMBo methods demonstrate this ability and affirm that these
methods outperform other known techniques.

\bibliographystyle{IEEEtran}
\bibliography{IEEEabrv,moshiko_bib}

\end{document}

%% file: ReMBo_alg.tex
\begin{algorithm}[h]
\caption{ReMBo (Reduce MMV and Boost)}\label{Alg1}
\begin{algorithmic}[1]
\REQUIRE $\b(Y),\b(A)$\\
\textbf{Control Parameters:} $K$, $\epsilon$, $\textsf{MaxIters}$,
$\mathcal{S}$, $\mathcal{P}$

\ENSURE $\b(\hat{X})$, $\hat{S}$, \textsf{flag}

\STATE Set \textsf{iter}$= 1$

\STATE Set \textsf{flag=false}

\WHILE{(\textsf{iter} $\leq$ \textsf{MaxIters}) and
(\textsf{flag} is \textsf{false})}

\STATE draw a random vector $\b(a)$ of length $d$ according to
$\mathcal{P}$. \label{StepRed1}

\STATE $\b(y)=\b(Y)\b(a)$\label{Step_y}

\STATE Solve $\b(y)=\b(A)\b(x)$ using SMV technique $\mathcal{S}$.
Denote the solution $\b(\hat{x})$. \label{StepSMV}

\STATE $\hat{S}=I(\b(\hat{x}))$ \label{StepRed2}

\IF{($|\hat{S}|\leq K$) and
($\|\b(y)-\b(A)\b(\hat{x})\|_2\leq\epsilon$)} \label{StepExamine}

\STATE \textsf{flag=true}

\ELSE \STATE \textsf{flag=flase} \ENDIF

\STATE Construct $\b(\hat{X})$ using $\hat{S}$ and
(\ref{Reconstruct1})-(\ref{Reconstruct2})

\STATE \textsf{iter}=\textsf{iter}$+1$

\ENDWHILE

%
%
%

\RETURN $\b(\hat{X})$, $\hat{S}$, \textsf{flag}

\end{algorithmic}
\end{algorithm}

%% file: list_techniques.tex
\begin{table*} \caption{Sub-Optimal Techniques}
\begin{center}
\begin{tabular}[C]{|c|l|l|l|}   \hline
Model & Tag & Formal Description & Type \\
\hline


\multirow{3}{*}{SMV}

& BP & Basis Pursuit,
(\ref{SMVP0}) with objective $\|\b(x)\|_1$, see \cite{Donoho},\cite{CandesRobust} & Convex relaxation \\

& OMP & Orthogonal Matching Pursuit, see \cite{Cotter} &
Greedy \\

& FOCUSS & FOcal Underdetermined System Solver, see \cite{Gorod}
& Greedy \\ \hline


\multirow{7}{*}{MMV}

& M-BP-$\ell_1$ \cite{Chen}& (\ref{MMVP0}) with objective
$\|\m(R)_{\ell_1}(\b(X))\|_1$ & Convex relaxation \\

& M-BP-$\ell_\infty$ \cite{TroppII} & (\ref{MMVP0}) with objective
$\|\m(R)_{\ell_\infty}(\b(X))\|_1$ & Convex relaxation \\

& M-OMP  & MMV version of OMP, see \cite{Cotter} & Greedy \\

& M-FOCUSS & MMV version of FOCUSS, see \cite{Cotter} & Greedy \\

\cline{2-4}

& ReMBo-BP & ReMBo with $\m(S)=$BP & Convex relaxation \\

& ReMBo-OMP & ReMBo with $\m(S)=$OMP & Greedy\\


\hline
\end{tabular}
\label{TableTechniques}
\end{center}
\end{table*}